\documentclass[12pt]{amsart}
\usepackage[margin = 1in]{geometry}
\newcommand{\norm}[1]{\lVert#1\rVert}
\newtheorem{thm}{Theorem}
\theoremstyle{definition}

\newtheorem{asmp}{Assumption}
\newtheorem{exm}{Example}
\begin{document}
\author{Andrey Sarantsev}
\title[Power Utility of Absolute and Relative Wealth]{Optimal Portfolio with Power Utility\\ of Absolute and Relative Wealth}
\address{Department of Mathematics and Statistics, University of Nevada, Reno} 
\email{asarantsev@unr.edu}

\begin{abstract}
Portfolio managers often evaluate performance relative to benchmark, usually taken to be the Standard \& Poor 500 stock index fund. This relative portfolio wealth is defined as the absolute portfolio wealth divided by wealth from investing in the benchmark (including reinvested dividends). The classic Merton problem for portfolio optimization considers absolute portfolio wealth. We combine absolute and relative wealth in our new utility function. We also consider the case of multiple benchmarks. To both absolute and relative wealth, we apply power utility functions, possibly with different exponents. We obtain an explicit solution and compare it to the classic Merton solution. We apply our results to the Capital Asset Pricing Model setting.
\end{abstract}

\maketitle

\thispagestyle{empty}

\section{Introduction}

\subsection{Portfolio utility optimization} Take a market model with finitely many risky assets and the risk-free asset. Each risky asset has wealth process (including reinvested dividends) evolving according to geometric Brownian motion. These Brownian motions might be correlated. The risk-free asset has wealth which is a deterministic exponential function. We consider a portfolio of these risky and risk-free assets. A central problem in quantitative finance is to maximize return and minimize risk of the wealth process $V(t)$ of this portfolio. 

One way to do this is to apply a concave utility function $U$ to the wealth at certain fixed time. Concavity implies diminishing marginal utility, corresponding to risk aversion. We choose portfolio weights to maximize expected utility of future wealth $\mathbb E U(V(T))$. Take one particular choice of utility: power function (or as its limit, the logarithmic function):
\begin{equation}
\label{eq:power}
U(x) = U_{\gamma}(x) := 
\begin{cases}
x^{\gamma}/\gamma,\, \gamma < 1,\, \gamma \ne 0;\\
\ln x,\, \gamma = 0.
\end{cases}
\end{equation}
For this function, the optimal portfolio has constant weights, obtained in the classic article \cite{Merton1, Merton2}. A solution for the general concave utility function is given in \cite{3people}. This power utility function has the property of {\it constant relative risk aversion}: $-(xU''(x))/U'(x) = 1 - \gamma$. For the general concave utility function, the solution is in \cite{3people}, and it is more complicated. There are two methods for solving these portfolio optimization problems. The first is the Girsanov change of measure, discussed in detail in \cite{3people}, and briefly in the textbook \cite[Exercise 7.5]{Oksendal}. The second is the Hamilton-Jacobi-Bellman equation, described briefly in the textbook \cite[Chapter 11]{Oksendal} and in much more detail in the monograph \cite{Krylov}. See also the classic book \cite[Section 5.8]{KS}, which includes both approaches. 

\subsection{Capital Asset Pricing Model} Recall that the equity premium is the difference between portfolio total returns and risk-free returns (usually taken to be short-term Treasury bills). The $\beta$ measure, called by its Greek letter name {\it beta}, is defined as the regression coefficient for the equity premium of a portfolio versus the equity premium of the benchmark. Mathematically, if $P(t)$ is the equity premium of a portfolio and $P_0(t)$ is the equity premium for the benchmark at time period $t$ (say, month or year), then we can write
\begin{equation}
\label{eq:CAPM0}
P(t) = \beta P_0(t) + \varepsilon(t), \quad \mathbb E[\varepsilon(t)] = 0.
\end{equation}
According to the Capital Asset Pricing Model (CAPM), proposed in \cite{Sharpe, Lintner}, $\beta$ is the only relevant risk measure: {\it market exposure}. Every other risk can be diversified away and does not merit higher returns. Later in this article, we modify~\eqref{eq:CAPM0} for continuous time. In subsequent research, this theory was disproved by real-world stock market data, see \cite{Survey} and references therein. Other risk factors were found which do not correspond to market exposure. However, the $\beta$ is still widely used by academics, professionals, and common investors. 

Sometimes portfolio managers evaluate performance not in terms of absolute wealth, but in comparison with a certain benchmark or several benchmark. If $V_1(\cdot), \ldots, V_d(\cdot)$ are wealth processes for these benchmarks, then we take a new utility function. It is a product of  utility functions of an absolute and relative portfolio wealth terms: 
$$
V(T), \quad \frac{V(T)}{V_1(T)}, \ldots, \frac{V(T)}{V_d(T)}.
$$
All these utility functions are concave power functions, possibly with different exponents. Thus we take into account both absolute and relative wealth. 

We consider a system of multiple risky assets, one of which serves as the benchmark; and one risk-free asset. All risky assets have wealth processes governed by stochastic differential equations driven by (correlated) Brownian motions. The risk-free wealth is deterministic.

As a real-world example of our computation, one can think of the benchmark as an S\&P 500 large-cap index fund, and other risky assets as other stock or bond index funds (for example, S\&P 400 mid-cap, S\&P 600 small-cap, and international stock markets). The risk-free asset is short-term Treasury bills.  For the USA, this benchmark is usually the Standard \& Poor 500 index fund, which contains stocks 500 large U.S. companies. Or we could use as benchmark the classic 60/40 portfolio: 60\% in Standard \& Poor 500 and 40\% in short-term risk-free Treasury bills. Usually, in practice there is only one benchmark. But we might as well consider several benchmarks combined in our utility function. 

\subsection{Literature review} The topic of this article is related to {\it principal-agent framework}, when the owner of the capital ({\it the principal}) employs the wealth manager ({\it the agent}). This creates conflicts of interest. For example, if the agent is compensated by a constant salary, he does not have any incentives to improve his work. On the contrary, if the agent is compensated by the percentage of net income (but not net losses), he will make risky investments: He captures the upside but is not harmed by the downside. 

There exists a vast literature on the principal-agent problem. Without attempt to give an extensive survey, we mention the book \cite{Cvitanic} for continuous time and  \cite{Prendergast} for discrete time, and references therein; as well as the following two old articles. In the article \cite{Bhattacharya}, the principal-agent problem was considered in a one-step model, with both principal and agent having exponential (constant absolute risk aversion) utilities. In \cite{HM}, a multistep and its extension (a continuous-time model) was considered, with the agent controlling the drift. 

Another related area of research is active management. Is it worth its fees and costs, compared to passive benchmark tracking (index funds)? We measure performance by excess return compared to the benchmark, or the {\it tracking error}. Analysis in a mean-variance framework was done in \cite{Roll}. The article \cite{3R} introduces a utility based on absolute return, variance, and the tracking error. The article \cite{TrackingError1} deals with measuring the portfolio wealth against a benchmark with constraints on weights to prevent excessive concentration in one asset. The article \cite{TrackingError2} includes also constraints on $\beta$. Both these articles deal with a one-step (not dynamic) problem, within the classic Modern Portfolio Theory framework. Linear deviation functions are used in \cite{TrackingError3} which are preferred over square deviations because of linear performance fees of fund managers. The Value at Risk constraints are studied in \cite{Alexander}. Finally, \cite{Stucchi} investigates the relationships between different approaches wth various restrictions. Another unified framework is developed in \cite{All}. 

We would like to mention yet another possible application. There exist some passive strategies (small, value, momentum) which outperform the stock market in the long run; see a recent article \cite{5factor} and a textbook \cite[Section 7]{Ang}, with references therein. However, they can underperform the market in the short run, which can last for several years. The investor then faces a conflict between optimizing absolute and relative wealth. We see this in Example 1 of Section 2 below: The result (compared to the classic utility maximization of absolute wealth) simply investing less proportion of wealth in the benchmark, and the same proportion in the alternative asset; but higher proportion in the risk-free asset. 

\subsection{Organization and results of the article} We solve this portfolio optimization problem explicitly using the Girsanov theorem in Theorem~\ref{thm:terminal}. We compare our solution with the classic Merton solution. The result is that we decrease the portfolio weight corresponding to the benchmark, but leave the other portfolio weights intact. Next, in Theorems~\ref{thm:CAPM} and~\ref{thm:CAPM-new} we apply our results to a continuous-time CAPM. We impose a constraint $\beta = \beta_0$. In the CAPM, we consider two options. When we can invest in the benchmark, and we measure our performance relative to the benchmark, the constraint on $\beta$ implies that the optimal portfolio is the benchmark and the risk-free asset. When we cannot invest in the benchmark, then the optimal portfolio does not depend on benchmarks at all. Section 2 contains the setting of the model and all main statements, while proofs are deferred until Section 3.

\subsection{Suggestion for future research} A natural question is to extend results to more general concave utility function: first a univariate $\mathbb U : (0, \infty) \to \mathbb R$ applied to the absolute or relative wealth terms; then a multivariate concave function of all these absolute and relative wealth terms. As mentioned above, this problem for absolute utility case was solved in \cite{3people} and discussed also in \cite[Section 5.8]{KS}. A more modest goal would be to logarithmic utility functions, since it can be considered a limiting case for a power utility function.  

\section{Definitions and Results} 

\subsection{Model setting} We operate on a filtered probability space $(\Omega, \mathcal F, (\mathcal F_t)_{t \ge 0}, \mathbb P)$ with the filtration satisfying the usual conditions ({\it completeness:} for $t \ge 0$, if $B \subseteq A \in \mathcal F_t$ and $\mathbb P(A) = 0$, then $B \in \mathcal F_t$ and therefore $\mathbb P(B) = 0$; and {\it right-continuity}: $\mathcal F_t = \cap_{s > t}\mathcal F_s$). Fix an $N \ge 1$, and take an $N$-dimensional Brownian motion $\mathbf{W} = (W_1, \ldots, W_N)$. Each $W_i$ has the Gaussian independent increments property: for $0 \le s \le t$, $W_i(t) - W_i(s) \sim \mathcal N(0, t-s)$ is independent of $\mathcal F_s$, each $W_i$ has continuous trajectories, and $W_1, \ldots, W_N$ are independent. Next, take an $N$-dimensional vector function $\mathbf{g} : \mathbb R_+ \to \mathbb R^{N}$, and an $N\times N$-matrix-valued symmetric positive definite function $\mathbf{A}$ on $\mathbb R_+$. Take yet another real-valued function $r : \mathbb R_+ \to \mathbb R$. Take a system of stochastic differential equations for $\mathbf{S} = (\mathbf{S}(t),\, t \ge 0)$, $\mathbf{S}(t) = (S_1(t), \ldots, S_N(t))$:
$$
\frac{\mathrm{d}\mathbf{S}(t)}{\mathbf{S}(t)} = \left(\frac{\mathrm{d}S_1(t)}{S_1(t)}, \ldots, \frac{\mathrm{d}S_N(t)}{S_N(t)}\right) = \mathbf{g}(t)\,\mathrm{d}t + \mathbf{\Sigma}(t)\,\mathrm{d}\mathbf{W}(t).
$$
Initial conditions are $S_i(0) = 1$ for $i = 1, \ldots, N$. Here, $\mathbf{\Sigma}(t)$ is the matrix square root of $\mathbf{A}(t)$. Take another (ordinary) differential equation: $\mathrm{d}B(t) = r(t)B(t)\,\mathrm{d}t$, $B(0) = 1$. Then $S_i(t)$ is the total wealth process (including reinvested dividends or interest payments) of the $i$th risky asset, and $B(t)$ is the total wealth process of the risk-free asset. A {\it portfolio} is an $N$-dimensional adapted process $\pi(t) = (\pi_1(t), \ldots, \pi_N(t))$. The quantity $\pi_i(t)$ is the proportion of wealth at time $t$ invested in the $i$th risky asset. The rest, that is $1 - \pi_1(t) - \ldots - \pi_N(t)$, is invested in the risk-free asset. We allow portfolio weights to be negative ({\it short positions}). The {\it wealth process} $V  = (V(t),\, t \ge 0)$ corresponding to this portfolio satisfies the equation:
$$
\frac{\mathrm{d}V(t)}{V(t)} = \sum_{i=1}^N\pi_i(t)\frac{\mathrm{d}S_i(t)}{S_i(t)} + \left(1 - \sum_{i=1}^N\pi_i(t)\right)\frac{\mathrm{d}B(t)}{B(t)}.
$$
The initial condition is again $V(0) = 1$. Now we define the $k$ benchmark portfolios $\rho_1, \ldots, \rho_k$. The $j$th benchmark portfolio is given by $\rho_j = (\rho_{j1}(t), \ldots, \rho_{jN}(t))$ for $j = 1, \ldots, k$, and has wealth process $V_j$.  

\subsection{Terminal utility: the main result} Fix constants $\gamma, \gamma_1, \ldots, \gamma_k \in \mathbb R$. 

\begin{asmp} We have $\gamma \in (0, 1)$. 
\label{asmp:gamma}
\end{asmp}

Take the following utility function 
\begin{equation}
\label{eq:utility}
\mathcal U := V^{1 - \gamma}(T)V_1^{-\gamma_1}(T)\cdot\ldots\cdot V_k^{-\gamma_k}(T).
\end{equation}
We choose a portfolio $\pi$ to maximize expected utility $\mathbb E\, [\mathcal U]$. We can rewrite~\eqref{eq:utility} as the product of $k+1$ power (constant relative risk aversion) utility functions, one of the absolute portfolio wealth, other $k$ are of portfolio wealth relative to the benchmark $\rho_k$:
$$
\mathcal U = V^{1 - \gamma - \gamma_1 - \ldots - \gamma_k}(T)\cdot  \left(\frac{V(T)}{V_1(T)}\right)^{\gamma_1}\cdot\ldots\cdot\left(\frac{V(T)}{V_k(T)}\right)^{\gamma_k}.
$$
Each parameter $\gamma_i$ has the meaning of the importance of relative wealth for this utility. As an example, a large $\gamma_1$ means the investor cares a lot about performance relative to the first benchmark $V_1$. Theoretically, it is possible to have $\gamma_i < 0$, but we could not find economic meaning of this case. Recall that, by Assumption 1, we {\it always} have $\gamma \in (0, 1)$.  

Let us introduce some notation. For two vectors $\mathbf{a}, \mathbf{b}$ of the same dimension, $\mathbf{a}\cdot \mathbf{b}$ denotes their inner product. The Euclidean norm of a vector $\mathbf{a}$ is defined as $\norm{\mathbf{a}} = \left[\mathbf{a}\cdot\mathbf{a}\right]^{1/2}$. Vectors $\mathbf{0} = (0, \ldots, 0)$ and $\mathbf{1} = (1, \ldots, 1)$ contain only components 0 and 1, respectively. Denote $\mathbf{e} = (1, 0, \ldots, 0)$. The dimension of these special vectors is clear from the context.

\begin{asmp}
The functions $\mathbf{g}, r, \mathbf{A}$ are bounded.
\end{asmp}

\begin{asmp}
The function $\mathbf{A}$ is {\it uniformly elliptic:} There exist $\varepsilon > 0$ such that for $\mathbf{x} \in \mathbb R^{N}$ and $t \ge 0$ we have: $\mathbf{A}(t)\mathbf{x}\cdot \mathbf{x} \ge \varepsilon\norm{\mathbf{x}}^2$. 
\end{asmp}

\begin{thm} Under Assumptions 1--3, among bounded portfolios $\pi$ the expected utility $\mathbb E[\mathcal U]$ for $\mathcal U$ from~\eqref{eq:utility} is maximized for 
\begin{equation}
\label{eq:pi}
\pi(t) = \frac1{\gamma}\mathbf{A}^{-1}(t)\left[\mathbf{g}(t) - r(t)\mathbf{1}\right] - \frac{1}{\gamma}\sum\limits_{j=1}^k\gamma_j\rho_j(t).
\end{equation}
\label{thm:terminal}
\end{thm}

We compare this result with classic Merton problem, which we get for $\gamma_1 = \ldots = \gamma_k = 0$: Here we have additional terms $\gamma_j\rho_j(t)/\gamma$. Thus we invest less in the benchmark, but other portfolio weights are the same as for the classic Merton problem.

\begin{exm} One particular case is when we select one risky asset $V_1$ as the benchmark. Usually this is a large-cap U.S. index, such as Standard \& Poor 500 or Russell 1000. We have $k = 1$ and $\rho_{11} = 1$, $\rho_{12} = \ldots = \rho_{1N} = 0$. The equation~\eqref{eq:pi} becomes
$$
\pi(t) = \frac1{\gamma}\mathbf{A}^{-1}(t)\left[\mathbf{g}(t) - r(t)\mathbf{1}\right] - \frac{\gamma_1}{\gamma}\mathbf{e}.
$$
\end{exm}


\begin{exm}
More generally, take the benchmark to be constant-weighted portfolio of the first risky asset (a large-cap U.S. index) and the risk-free asset (short-term Treasury bills), with proportions $\theta$ and $1 - \theta$. A classic example is 60/40 portfolio which invests 60\% in S\&P 500 and 40\% in T-bills. Then we have: $\rho_{11} = \theta$, $\rho_{12} = \ldots = \rho_{1N} = 0$, and~\eqref{eq:pi} takes the form:
$$
\pi(t) = \frac1{\gamma}\mathbf{A}^{-1}(t)\left[\mathbf{g}(t) - r(t)\mathbf{1}\right] - \frac{\gamma_1\theta}{\gamma}\mathbf{e}.
$$
\end{exm}

\subsection{Capital Asset Pricing Model} As mentioned in the Introduction, the Capital Asset Pricing Model (CAPM) considers $\beta$ (market exposure) as the only risk measure. The {\it equity premium} $P$ of a risky asset is defined as the difference between the returns of this risky asset and the returns of the risk-free asset. The Capital Asset Pricing Model states that every risky asset has a constant risk measure $\beta$ such that the equity premium $P$ of this risky asset equals 
$P = \beta P_0 + \varepsilon$, where $P_0$ is the equty premium of the benchmark, and $\varepsilon$ is a residual, with zero mean. Then the returns of the risky asset can be expressed as a linear combination of the benchmark returns (with coefficient $\beta$), the risk-free returns (with coefficient $1 - \beta$), and a residual. In continuous time, the residual 
$\varepsilon$ is replaced by $\mathrm{d}W_i(t)$, where $W_i$ is a Brownian motion. Thus,
\begin{equation}
\label{eq:CAPM}
\frac{\mathrm{d}S_i(t)}{S_i(t)} = \beta_i\frac{\mathrm{d}S_0(t)}{S_0(t)} + (1 - \beta_i)\,\frac{\mathrm{d}B(t)}{B(t)} + \mathrm{d}W_i(t),\quad i = 1, \ldots, N,
\end{equation}
where $W_1, \ldots, W_N$ form an $N$-dimensional Brownian motion with drift vector zero and covariance matrix $\mathbf{C}$; and $\beta_1, \ldots, \beta_N$ are risk measures for $N$ risky assets. As before,
\begin{equation}
\label{eq:bond-benchmark}
\mathrm{d}B(t) = rB(t)\,\mathrm{d}t,\quad \mathrm{d}S_0(t) = S_0(t)(\mu\,\mathrm{d}t + \sigma \,\mathrm{d}W_0(t)).
\end{equation}
Here, $W_0$ is the standard Brownian motion, independent of $\mathbf{W} = (W_1, \ldots, W_N)$, and $r < g$. The latter condition shows that the growth rate of the benchmark (a risky asset) is greater than that of the risk-free asset.  Combined,~\eqref{eq:CAPM} and~\eqref{eq:bond-benchmark} create a simple CAPM setting in continuous time. This is a particular case of the model in the previous subsection: $(S_0, S_1, \ldots, S_N)$. Here, the drift vector given by
\begin{equation}
\label{eq:drift-CAPM}
\mathbf{g}(t) = \mu\begin{bmatrix} 1 \\ \mathbf{b}\end{bmatrix} + r\begin{bmatrix}0 \\ \mathbf{1} - \mathbf{b}\end{bmatrix},\quad \mathbf{b} = (\beta_1, \ldots, \beta_N)^T,
\end{equation}
and the covariance matrix is given by the block-diagonal form:
\begin{equation}
\label{eq:A-CAPM}
\mathbf{A} = 
\begin{bmatrix}
\sigma^2 & \mathbf{b}\sigma^2\\
\mathbf{b}^T\sigma^2 & \mathbf{C} + \mathbf{b}\mathbf{b}^T\sigma^2
\end{bmatrix}
\end{equation}
We have $N+1$ risky assets instead of $N$. The portfolio in this case is defined as $\mathbf{\pi} = (\pi_0, \pi_1, \ldots,\pi_N)$, with wealth process defined as 
$$
\frac{\mathrm{d}V(t)}{V(t)} = \sum\limits_{i=0}^N\pi_k(t)\frac{\mathrm{d}S_k(t)}{S_k(t)} + (1 - \pi_0(t) - \pi_1(t) - \ldots - \pi_N(t))\frac{\mathrm{d}B(t)}{B(t)}.
$$
The risk measure $\beta$ of a portfolio $\mathbf{\pi} = (\pi_0, \pi_1,\ldots,\pi_N)$ is computed as the linear combination of $\beta$s of individual risky assets: $\pi_0(t) + \beta_1\pi_1(t) + \ldots + \beta_N\pi_N(t)$. Fix a value $\beta_0 > 0$ and impose the condition
\begin{equation}
\label{eq:fixed-beta}
\pi_0(t) + \beta_1\pi_1(t) + \ldots + \beta_N\pi_N(t) = \beta_0.
\end{equation}
Now let us maximize expected terminal utility~\eqref{eq:utility} under this restriction~\eqref{eq:fixed-beta}. Denote 
\begin{equation}
\label{eq:v}
\begin{bmatrix}
v_0 \\ \mathbf{v} 
\end{bmatrix}
= \sum\limits_{j=1}^k\gamma_jA\rho_j.
\end{equation}

\begin{thm}
Among bounded portfolios, the one which maximizes $\mathbb E[\mathcal U]$ for $\mathcal U$ from~\eqref{eq:utility} given~\eqref{eq:fixed-beta} is equal to 
\begin{equation}
\label{eq:portfolio-CAPM}
\begin{bmatrix} 
\pi_1(t) & \ldots & \pi_N(t) 
\end{bmatrix}
= \sigma^2\mathbf{C}^{-1}(\mathbf{v} - v_0\mathbf{b}). 
\end{equation}
\label{thm:CAPM}
\end{thm}

\begin{exm}
Consider the case of one benchmark portfolio ($k = 1$), with $\rho_{10} = \theta$ and $\rho_{11} = \ldots = \rho_{1N} = 0$. This portfolio invests the fixed proportion $\theta$ into the benchmark $S_0$ and the rest $1 - \theta$ in the risk-free asset. Then $\rho_1 = \theta\mathbf{e}$, and the vector from~\eqref{eq:v} becomes 
$$
\gamma_1\theta
\begin{bmatrix}
\sigma^2 & \sigma^2\mathbf{b} \\ \sigma^2\mathbf{b}^T & \sigma^2\mathbf{b}\mathbf{b}^T + \mathbf{C} 
\end{bmatrix}
\begin{bmatrix} 1 \\ \mathbf{0} \end{bmatrix} = 
\gamma_1\theta\sigma^2\begin{bmatrix} 1 \\ \mathbf{b} \end{bmatrix}
$$
Therefore, $v_0 = \gamma_1\theta\sigma^2$ and $\mathbf{v} = \gamma_1\theta\sigma^2\mathbf{v}$. 
Among bounded portfolios, the one which maximizes 
$\mathbb E[\mathcal U]$ for $\mathcal U$ from~\eqref{eq:utility} given~\eqref{eq:fixed-beta} is given by $\pi_1(t) = \ldots = \pi_N(t) = 0$, $\pi_0(t) = \beta_0$. 

Amazingly, the portfolio weights corresponding to risky assets other than the benchmark are simply $0$. The optimal portfolio with restricted $\beta$ in the CAPM benchmark does not invest in any risky assets, except the benchmark. It does not depend on $\rho_1, \ldots, \rho_k$, and on growth rate and volatility $\mu, \sigma$. The same is true if we did not have any benchmark portfolios, only the classic power utility, as in the Merton problem. The restriction on $\beta$ surprisingly simplifies the answer. 
\end{exm}

Assume now we cannot invest directly in a benchmark. Then a portfolio will be defined as a process $(\pi_1(t), \ldots, \pi_N(t))$. Here as before, $\pi_i(t)$ is the proportion invested in the $i$th risky asset, and $1 - \pi_1(t) - \ldots - \pi_N(t)$ is the proportion invested in the risk-free asset. We can similarly compute the drift vector and the covariance matrix
\begin{equation}
\label{eq:drift-covar-new}
\mathbf{g} = (\mu - r)\mathbf{b} + r\mathbf{1},\quad \mathbf{A} = \mathbf{C} + \sigma^2\mathbf{b}\mathbf{b}^T.
\end{equation}
The constraint on $\beta$ is given by $\mathbf{b}\cdot \mathbf{p} = \beta_0$. The solution to the optimal investment problem in this case is more complicated than in Theorem~\ref{thm:CAPM}.

\begin{thm}
\label{thm:CAPM-new}
Among bounded portfolios, the one which maximizes $\mathbb E[\mathcal U]$ for $\mathcal U$ from~\eqref{eq:utility} given $\mathbf{b}\cdot \mathbf{\pi}(t) = \beta_0$ is equal to 
\begin{equation}
\label{eq:portfolio-CAPM-new}
\begin{bmatrix} 
\pi_1(t) & \ldots & \pi_N(t) 
\end{bmatrix}
= \frac{\beta_0\mathbf{C}^{-1}\mathbf{b}}{\mathbf{b}^T\mathbf{C}^{-1}\mathbf{b}}. 
\end{equation}
\end{thm}

Here we do not have dependence upon the benchmarks. Just like for the classic Merton problem, with $\gamma_i = 0$ for all $i$, we have the same answer. 

\section{Proofs} 

\subsection{Proof of Theorem~\ref{thm:terminal}} Assume without loss of generality, that benchmark portfolios $\rho_1, \ldots, \rho_k$, and functions $r, \mathbf{g}, \mathbf{A}$ are constant (do not depend on $t$). This will slightly simplify the notation, but will not change the proof in any essential  way. Consider the wealth process:
\begin{align*}
\frac{\mathrm{d}V(t)}{V(t)} &= \sum\limits_{k=1}^N\pi_k(t)\frac{\mathrm{d}S_k(t)}{S_k(t)} + (1 - \pi_1(t) - \ldots - \pi_N(t))\frac{\mathrm{d}B(t)}{B(t)} \\ & = \left[\pi_1(t)g_1 + \ldots + \pi_N(t)g_N + (1 - \pi_1(t) - \ldots - \pi_N(t))r\right]\,\mathrm{d}t + \pi(t)\cdot \mathbf{\Sigma}(t)\,\mathrm{d}\mathbf{W}(t).
\end{align*}
Applying It\^o's formula, we get:
\begin{align}
\label{eq:ito}
\begin{split}
\mathrm{d}&\ln V(t) = \frac{\mathrm{d}V(t)}{V(t)} - \frac{\mathrm{d}\langle V\rangle_t}{2V^2(t)} = \pi(t)\cdot \mathbf{\Sigma}(t)\,\mathrm{d}\mathbf{W}(t)\\ & +  \left[\pi_1(t)g_1 + \ldots + \pi_N(t)g_N + (1 - \pi_1(t) - \ldots - \pi_N(t))r - \mathbf{A}(t)\pi(t)\cdot \pi(t)/2\right]\,\mathrm{d}t.
\end{split}
\end{align}
Similarly, applying this to the benchmark portfolio $\rho_j$ instead of $\pi$, we get:
\begin{align}
\label{eq:ito-benchmark}
\begin{split}
\mathrm{d}&\ln V_j(t) = \rho_j(t)\cdot \mathbf{\Sigma}(t)\,\mathrm{d}\mathbf{W}(t) \\ & + \left[\rho_{j1}(t)g_1 + \ldots + \rho_{jN}(t)g_N + (1 - \rho_{j1}(t) - \ldots - \rho_{jN}(t))r(t)\right]\,\mathrm{d}t \\ & -  \mathbf{A}(t)\rho_j(t)\cdot \rho_j(t)/2\,\mathrm{d}t
\end{split}
\end{align}
Next, we can rewrite the utility as
$$
\mathcal U = \exp((1 - \gamma)\ln V(T) - \gamma_1\ln V_1(T) - \ldots - \gamma_k\ln V_k(T)).
$$
Combining the stochastic equations~\eqref{eq:ito} and~\eqref{eq:ito-benchmark}, we get:
\begin{align}
\label{eq:ito-log}
\mathrm{d}\left[(1 - \gamma)\ln V(T) - \gamma_1\ln V_1(T) - \ldots - \gamma_k\ln V_k(T)\right] = F(\pi(t))\,\mathrm{d}t + \mathrm{d}\mathcal M(t),  
\end{align}
Here the function $F : \mathbb R^{N} \to \mathbb R$ is defined by 
\begin{align*}
F(\mathbf{p}) &= (1 - \gamma) \left[\mathbf{p} \cdot \mathbf{g} + (1 - p_1 - \ldots - p_N)r - \frac12\mathbf{A}\mathbf{p}\cdot \mathbf{p}\right] \\ & - \sum\limits_{j=1}^k\gamma_j\left[\rho_j\cdot \mathbf{g} + (1 - \mathbf{1}\cdot \rho_j)r - \frac12 \mathbf{A}\rho_j\cdot \rho_j\right].
\end{align*}
In addition, $\mathcal M = (\mathcal M(t),\, t \ge 0)$ is a martingale: 
$$
\mathrm{d}\mathcal M(t) = -\sum\limits_{j=1}^k\gamma_j\sum\limits_{i=1}^N\rho_{ji}(t)\mathrm{d}W_i(t) + (1 - \gamma)\sum\limits_{i=1}^N\pi_i\,\mathrm{d}W_i(t).
$$
The quadratic variation of $\mathcal M$ is given by
$$
\mathrm{d}\langle \mathcal M\rangle_t = G(\pi(t))\,\mathrm{d}t,\quad G(\mathbf{p}) := \mathbf{A}\mathbf{v}(\mathbf{p})\cdot \mathbf{v}(\mathbf{p}), \quad \mathbf{v}(\mathbf{p}) = (1 - \gamma) \mathbf{p} - \gamma_1\rho_1 - \ldots - \gamma_k\rho_k.
$$
Since $\pi(t)$ is bounded, and $G$ is continuous, the Novikov condition holds: $\mathbb E\exp\left(\langle M\rangle_T/2\right) < \infty$. Thus by the Girsanov theorem, we get: 
$$
\mathbb E[\mathcal U] = \mathbb E^*\exp\left[\int_0^T\left(F(\pi(t))\, + \frac12G(\pi(t))\right)\,\mathrm{d}t\right].
$$
Here, $\mathbb E^*$ is the expectation with respect to the new probability measure $\mathbb P^*$, absolutely continuous with respect to the original measure $\mathbb P$, with the density (Radon-Nikodym derivative):
$$
\frac{\left.\mathrm{d}\mathbb P^*\right|_{\mathcal F_T}}{\left.\mathrm{d}\mathbb{P}\right|_{\mathcal F_T}} = \exp\left[\mathcal M(T) - \frac12\langle \mathcal M\rangle_T\right].
$$
The following function is quadratic:
\begin{equation}
\label{eq:H}
H := F + G/2 : \mathbb R^{N} \to \mathbb R.
\end{equation}
Take the gradient $\nabla H$:
\begin{align}
\label{eq:gradient}
\begin{split}
\nabla H &= (1 - \gamma)(\mathbf{g} - r\mathbf{1} - \mathbf{A}\mathbf{p}) + (1 - \gamma)\mathbf{A}((1 - \gamma) \mathbf{p} - \gamma_1\rho_1 - \ldots - \gamma_k\rho_k) \\ & = -\gamma(1 - \gamma) \mathbf{A}\mathbf{p} + (1 - \gamma)(\mathbf{g} - r\mathbf{1} - \mathbf{A}(\gamma_1\rho_1 + \ldots + \gamma_k\rho_k)).
\end{split}
\end{align}
From~\eqref{eq:gradient}, the Hessian (second derivative matrix) of $H$ is equal to $-\gamma(1 - \gamma)\mathbf{A}$. This Hessian is negative definite if $\gamma \in (0, 1)$, which is an assumption of Theorem~\ref{thm:terminal}. Thus, to maximize $H$ from~\eqref{eq:H}, solve the vector equation $\nabla H = 0$. The unique solution is 
$$
\mathbf{p} = -\frac{1}{\gamma}\sum\limits_{j=1}^k\gamma_j\rho_j + \frac1{\gamma}\mathbf{A}^{-1}\left(\mathbf{g} - r\mathbf{1}\right).
$$
This completes the proof of the main result. 

\subsection{Proof of Theorem~\ref{thm:CAPM}} Following the proof of Theorem~\ref{thm:terminal}, we again maximize the function $H$ from~\eqref{eq:H} over $p = (q_0, \mathbf{q})$ with $q_0 \in \mathbb R$ and $\mathbf{q} \in \mathbb R^N$ which satisfy the following constraint:
\begin{equation}
\label{eq:constraint}
q_0 + \mathbf{b}\cdot \mathbf{q} = \beta_0,\quad \mathbf{b} = (\beta_1, \ldots, \beta_N).
\end{equation}
The fraction invested in risk-free assets is $p_0 = 1 - q_0 - \mathbf{q}\cdot\mathbf{1} = 1 - \beta_0 + \mathbf{b}\cdot \mathbf{q} - \mathbf{q}\cdot\mathbf{1}$. We use Lagrange optimization. The gradient of the left-hand side in the linear condition~\eqref{eq:constraint} is equal to $\begin{bmatrix} 1 \\ \mathbf{b} \end{bmatrix}$ and thus the Lagrange equation is: for some $\lambda \in \mathbb R$,
\begin{equation}
\label{eq:gradient-lagrange}
\nabla H + \lambda\begin{bmatrix} 1 \\ \mathbf{b} \end{bmatrix} = \begin{bmatrix} 0 \\ \mathbf{0} \end{bmatrix} \in \mathbb R^{N+1}.
\end{equation}
We need to solve this equation~\eqref{eq:gradient-lagrange} together with~\eqref{eq:constraint}. This is a system of $N+2$ equations with $N+2$ unknown variables $q_0, \mathbf{q}, \lambda$. Plug $p = (q_0, \mathbf{q})$ into~\eqref{eq:gradient} and use expression for the drift vector and the covariance matrix from~\eqref{eq:drift-CAPM} and~\eqref{eq:A-CAPM}. Using the equation~\eqref{eq:constraint}, we get:
$$
\mathbf{A}\mathbf{p} = \begin{bmatrix} 
\sigma^2 & \mathbf{b}\sigma^2\\
\mathbf{b}^T\sigma^2 & \mathbf{C} + \mathbf{b}\mathbf{b}^T\sigma^2
\end{bmatrix}
\begin{bmatrix}
q_0 \\ \mathbf{q} 
\end{bmatrix} = 
\begin{bmatrix}
\sigma^2(q_0 + \mathbf{b}\cdot \mathbf{q}) \\ 
\sigma^2b(q_0 + \mathbf{b}\cdot \mathbf{q}) + \sigma^2\mathbf{C}\mathbf{q}
\end{bmatrix}
= \sigma^2\beta_0\begin{bmatrix} 1 \\ \mathbf{b}\end{bmatrix} + \sigma^2\begin{bmatrix} 0 \\ \mathbf{C}\mathbf{q} \end{bmatrix}
$$
Plugging this result and~\eqref{eq:drift-CAPM} into~\eqref{eq:gradient}, we get:
\begin{equation}
\mathbf{g} - r\mathbf{1} = \mu\begin{bmatrix}1 \\ \mathbf{b} \end{bmatrix} + r\begin{bmatrix}0 \\ \mathbf{1} - \mathbf{b}\end{bmatrix} - r\begin{bmatrix} 1 \\ \mathbf{1} \end{bmatrix} = (\mu - r)\begin{bmatrix} 1 \\ \mathbf{b}\end{bmatrix}
\end{equation}
The equation~\eqref{eq:gradient} becomes
\begin{equation}
\label{eq:gradient-CAPM}
\frac1{1 - \gamma}\nabla H = 
(\mu - r)\begin{bmatrix} 1 \\ \mathbf{b}\end{bmatrix} - \begin{bmatrix}v_0 \\ \mathbf{v}\end{bmatrix} - \gamma\mathbf{A}\mathbf{p}
\end{equation}
Plugging~\eqref{eq:A-CAPM} into~\eqref{eq:gradient-CAPM}, we get: 
$$
\frac1{1 - \gamma}\nabla H = (\mu - r)\begin{bmatrix} 1 \\ \mathbf{b}\end{bmatrix}  - \begin{bmatrix}v_0 \\ \mathbf{v}\end{bmatrix}  - \gamma\sigma^2\beta_0\begin{bmatrix} 1 \\ \mathbf{b}\end{bmatrix} - \gamma\sigma^2\begin{bmatrix} 0 \\ \mathbf{C}\mathbf{q} \end{bmatrix}
$$
Changing without loss of generality $\lambda$ to $\lambda/(1 - \gamma)$ in~\eqref{eq:gradient-lagrange}, and plugging there, we get:
\begin{equation}
\label{eq:main-eq}
(\mu - r - \gamma\beta_0\sigma^2 + \lambda)
\begin{bmatrix} 1 \\ \mathbf{b}\end{bmatrix} - \begin{bmatrix}v_0 \\ \mathbf{v} \end{bmatrix} +  \sigma^2\begin{bmatrix} 0 \\ \mathbf{C}\mathbf{q}\end{bmatrix} = 
\begin{bmatrix} 0 \\ \mathbf{0} \end{bmatrix}
\end{equation}
Comparing the first components in the left-hand and right-hand sides in~\eqref{eq:main-eq}, we get: $\mu - r - \gamma\beta_0\sigma^2 + \lambda = v_0$. Comparing the other $N$ components in~\eqref{eq:main-eq}, we get:
$$
(\mu - r - \gamma\beta_0\sigma^2 + \lambda)\mathbf{b} + \sigma^2\mathbf{C}\mathbf{q} = \mathbf{v},
$$
which implies $v_0\mathbf{b} + \sigma^2\mathbf{C}\mathbf{q} = \mathbf{v}$. The rest of the proof is trivial. 

\subsection{Proof of Theorem~\ref{thm:CAPM-new}} Similarly to the proof of Theorem~\ref{thm:CAPM}, we solve the quadratic programming problem: maximize $H(\mathbf{p})$ given $\mathbf{p}\cdot\mathbf{b} = \beta_0$. Applying the Lagrange optimization method, we get: for some constant $\lambda$, 
\begin{equation}
\label{eq:Lag}
\frac{\nabla H}{1 - \gamma} + \lambda \nabla(\mathbf{p}\cdot \mathbf{b} - \beta_0) = 0.
\end{equation}
Computing $\nabla H$ as in~\eqref{eq:gradient}, and using expressions from~\eqref{eq:drift-covar-new}, we get:
$$
\frac{\nabla H}{1 - \gamma} = -\gamma (\mathbf{C} + \sigma^2\mathbf{b}\mathbf{b}^T)\mathbf{p} + (\mu - r)\mathbf{b} - \sum\limits_{i=1}^k\gamma_i(\mathbf{C} + \sigma^2\mathbf{b}\mathbf{b}^T)\rho_i.
$$  
By~\eqref{eq:Lag}, this must be proportional to $\mathbf{b}$. Note that the terms with $\mathbf{b}\mathbf{b}^T$ are already proportional to $\mathbf{b}$. Therefore, for some real constant $\lambda'$ (of course, dependent on $\lambda$), we have:
$$
\mathbf{C}\mathbf{p} = -\sum\limits_{i=1}^k\frac{\gamma_i}{\gamma}\mathbf{C}\rho_i + \lambda' \mathbf{b}. 
$$
Multiplying by $\mathbf{C}^{-1}$, we get:
\begin{equation}
\label{eq:final}
\mathbf{p} = -\sum\limits_{i=1}^k\frac{\gamma_i}{\gamma}\rho_i + \lambda' \mathbf{C}^{-1}\mathbf{b}. 
\end{equation}
To satisfy $\mathbf{p}\cdot\mathbf{b} = \beta_0$, we plug~\eqref{eq:final} into it and solve for $\lambda'$. This gives us the result
$$
\mathbf{p} = -\sum\limits_{i=1}^k\frac{\gamma_i}{\gamma}\rho_i + \frac{1}{\mathbf{b}^T\mathbf{C}^{-1}\mathbf{b}}\left[\beta_0 + \sum\limits_{i=1}^k\frac{\gamma_i}{\gamma}\rho_i\cdot\mathbf{b}\right]\mathbf{C}^{-1}\mathbf{b}
$$
which after cancellation gives us~\eqref{eq:portfolio-CAPM-new}.

\section*{Acknowledgements}

The author is thankful to the Department of Mathematics and Statistics, University of Nevada, Reno, for support and friendly environment for applied research.

\bibliographystyle{plain}

\end{document}